\documentclass[aps,prd,reprint,showpacs,floatfix,longbibliography,nofootinbib,superscriptaddress]{revtex4-1}

\usepackage{tikz,graphicx,booktabs,multirow,array,microtype,mathtools,float}

\usepackage{centernot}
\newcommand{\fsl}[1]{{\centernot{#1}}} 

\graphicspath{{figures/}}
\usepackage{enumerate}   
\extrafloats{200}
\usepackage[colorlinks=true,backref=false, linktocpage=true,
citecolor=blue,
urlcolor=blue,linkcolor=blue,
pdfpagemode=UseOutlines]{hyperref}
\newcolumntype{C}{>{$}c<{$}}
\usepackage{dcolumn}

\usepackage{anyfontsize}
\usepackage[shortlabels]{enumitem}

\usepackage{amsfonts}
\AtBeginDocument{
\heavyrulewidth=.08em
\lightrulewidth=.05em
\cmidrulewidth=.03em
\belowrulesep=.65ex
\belowbottomsep=0pt
\aboverulesep=.4ex
\abovetopsep=0pt
\cmidrulesep=\doublerulesep
\cmidrulekern=.5em
\defaultaddspace=.5em
}

\def\Tr {\mathop{\hbox{Tr}}}

\newcommand{\langl}{\begin{picture}(4.5,7)
\put(1.1,2.5){\rotatebox{60}{\line(1,0){5.5}}}
\put(1.1,2.5){\rotatebox{300}{\line(1,0){5.5}}}
\end{picture}}
\newcommand{\rangl}{\begin{picture}(4.5,7)
\put(.9,2.5){\rotatebox{120}{\line(1,0){5.5}}}
\put(.9,2.5){\rotatebox{240}{\line(1,0){5.5}}}
\end{picture}}

\usepackage{amsmath} 
\usepackage{dsfont}  
\usepackage{bm}      

\def\iden{\mathds{1}}

\def\beq{\begin{equation}}
\def\eeq{\end{equation}}
\def\beqs#1\eeqs{\beq\begin{split} #1 \end{split}\eeq}

\long\def\comment#1{}

\newcommand{\comm}[1]{{\color{red}{\texttt{\color{red}\scriptsize[#1]}}}}

\makeatletter 
    
\renewcommand\onecolumngrid{
\do@columngrid{one}{\@ne}%
\def\set@footnotewidth{\onecolumngrid}
\def\footnoterule{\kern-6pt\hrule width 1.5in\kern6pt}%
}

\renewcommand\twocolumngrid{
        \def\footnoterule{
        \dimen@\skip\footins\divide\dimen@\thr@@
        \kern-\dimen@\hrule width.5in\kern\dimen@}
        \do@columngrid{mlt}{\tw@}
}%

\makeatother    

\begin{document}

\title{
Magnetic polarizability of a charged pion  from four-point functions in lattice QCD 
}
%
\author{Frank~X.~Lee}
\email{fxlee@gwu.edu}
\affiliation{Physics Department, The George Washington University, Washington, DC 20052, USA}
\author{Walter Wilcox}
\email{walter\_wilcox@baylor.edu}
\affiliation{Department of Physics, Baylor University, Waco, Texas 76798, USA}
\author{Andrei~Alexandru}
\email{aalexan@gwu.edu}
\affiliation{Physics Department, The George Washington University, Washington, DC 20052, USA}
\affiliation{Department of Physics, University of Maryland, College Park, MD 20742, USA}
\author{Chris~Culver}
\email{c.culver@liverpool.ac.uk}
\affiliation{Department of Mathematical Sciences, University of Liverpool, Liverpool L69 7ZL, United Kingdom}
%
%

\begin{abstract}
Electromagnetic dipole polarizabilities are fundamental properties of a hadron that represent its resistance to deformation under external fields. For a charged hadron, the presence of acceleration and  Landau levels  complicates the isolation of its deformation energy in the conventional background field method. In this work, we explore a general method based on four-point functions in lattice QCD that takes into account all  photon, quark and gluon interactions. 
 The electric polarizability ($\alpha_E$) has been determined from the method in a previous proof-of-principle simulation.
Here we focus on the magnetic polarizability ($\beta_M$) using  the same quenched Wilson action on a $24^3\times 48$ lattice at $\beta=6.0$ 
with pion mass from 1100 to 370 MeV.  The results from the connected diagrams show a large cancellation between the elastic and inelastic contributions, leading to a relatively small and negative value for $\beta_M$ consistent with chiral perturbation theory. We also discuss the mechanism for  $\alpha_E+\beta_M$ from combining the two studies.

\end{abstract}
\maketitle

\twocolumngrid

\section{Introduction}
\label{sec:intro} 

Understanding electromagnetic polarizabilities has been a long-term goal of lattice QCD. 
The standard approach is the background field method which introduces classical static electromagnetic fields to interact with quarks in QCD~\cite{PhysRevLett.49.1076,MARTINELLI1988865,Fiebig:1988en,Burkardt_1996,Alexandru:2008sj, Lee:2005ds, Lee:2005dq, Engelhardt:2007ub, Tiburzi:2008ma, Detmold_2006,Detmold_2009,Detmold:2009fr, Detmold:2010ts,Freeman:2013eta, Lujan:2014kia, Freeman:2014kka, Lujan:2016ffj,Primer_2014,Luschevskaya_2015,Luschevskaya:2015cko,Luschevskaya:2016epp,Chang:2015qxa,Parreno:2016fwu,Bali:2017ian,Bruckmann:2017pft,Deshmukh:2017ciw,Bignell_2018,Bignell_2019,Bignell_2020,Bignell:2020xkf,He:2021eha,niyazi2021charged}.  
The appeal of the method lies in its simplicity: only two-point correlation  functions are needed to measure the small energy shift with or without the external field, which amounts to a standard calculation of  a hadron's  mass. The energy shift linear in the applied field is related to dipole moments, and the quadratic shift to polarizabilities. The method is fairly robust and has been widely applied to neutral hadrons.

When it comes to charged hadrons, however, the method is faced with new challenges.
The reason is rather rudimentary: a charged particle accelerates in an electric field and exhibits Landau levels in a magnetic field.  
Such collective motion of the hadron is unrelated to moments and polarizabilities and must be disentangled from the total energy shift in order to isolate the deformation energy on which the polarizabilities are defined. 
The traditional method of extracting ground state energy at large times breaks down since  the two-point function no longer has a single-exponential behavior.
Special techniques have to be developed to analyze such functions. Nonetheless, progress has been made.
For electric field,  a continuum relativistic propagator for a charged scalar is used to demonstrate how to fit such lattice data for charged pions and kaons~\cite{Detmold_2006,Detmold_2009}. It is improved upon by an effective propagator exactly matching the lattice being used to generate the lattice QCD data~\cite{niyazi2021charged}. Furthermore, spatial and time profiles $G(x,t)$ under Dirichlet boundary conditions with both real and imaginary parts are used to capture the interactions while maintaining gauge invariance in the background field. 
For magnetic field, various techniques have been tried to deal with Landau levels, from direct fitting forms~\cite{Primer_2014,Luschevskaya_2015,Luschevskaya:2015cko,Luschevskaya:2016epp,Chang:2015qxa,Parreno:2016fwu}, 
to formal studies of Dirac operator~\cite{Bali:2017ian,Bruckmann:2017pft}, to a novel Laplacian-mode projection technique at the quark propagator level~\cite{Bignell_2018,Bignell_2019,Bignell_2020,Bignell:2020xkf,He:2021eha}.

Here we advocate  an alternative approach based on four-point functions in lattice QCD. 
Instead of background fields,  electromagnetic currents couple to quark fields. All photon-quark, quark-quark, and gluon-quark interactions are included.
It is a general approach that treats neutral and charged particles on equal footing.
The potential of using four-point functions to access polarizabilities has been investigated in the early days of lattice QCD~\cite{BURKARDT1995441,Andersen:1996qb,Wilcox:1996vx}. The effort was deemed too computationally demanding at the time and the results on limited lattices were inconclusive. Recently, there is  a renewed interest to revive such efforts, partly spurred by the challenges encountered in the background field method for charged particles.  A reexamination of the formalism in Ref.~\cite{Wilcox:1996vx} is carried out  in Ref.~\cite{Wilcox:2021rtt} in which 
new formulas are derived in momentum space for electric and magnetic polarizabilities of both charged pion and  proton.  
It is followed by a  proof-of-principle simulation for the electric polarizability of a charged pion~\cite{Lee:2023rmz}. 
In this work, we extend the calculation to magnetic polarizability using the same lattice parameters.
We note there exists other four-point function calculations on polarizabilities. Ref.\cite{Feng:2022rkr} employs a position-space formula for the Compton tensor to calculate charge pion electric polarizability near the physical point, along with  a  preliminary calculation on the proton~\cite{Wang:2021}. A comprehensive review on pion polarizabilities  from other theoretical approaches and experiment can be found in Ref.~\cite{Moinester:2022tba,Moinester:2019sew}.
We also note that although Refs.~\cite{Engelhardt:2007ub,Engelhardt:2011qq} are based on  the  background field
method, they are in fact four-point function calculations. A perturbative expansion in the background field at the action level is performed in which two vector current insertions couple the background field to the hadron correlation function, leading to the same diagrammatic structures as in this work.

In Sec.~\ref{sec:method} we outline the methodology to extract magnetic polarizability of a charged pion from four-point functions. 
In Sec.~\ref{sec:results} we  show our results from a proof-of-concept simulation. In particular, we discuss $\beta_M$ and its chiral extrapolation, $\alpha_E+\beta_M$, and comparison with ChPT.
In Sec.~\ref{sec:con} we give concluding remarks.
The four-point correlation functions needed in the simulation are given in the Appendix.

\section{Methodology}
\label{sec:method} 

In Ref.~\cite{Wilcox:2021rtt}, a formula is derived for electric polarizability of a charged pion,
\fontsize{9}{9}
\begin{align}
\alpha_E=  
 {\alpha \langl  r_E^2\rangl \over 3m_\pi}+\lim_{\bm q\to 0}{2\alpha \over \bm q^{\,2}} \int_{0}^\infty d t \bigg[Q_{44}(\bm q,t) -Q^{elas}_{44}(\bm q,t) \bigg],
 \label{eq:alpha}  
\end{align}
and for its magnetic polarizability,
\fontsize{8}{8}
\begin{align}
\beta_M=
- {\alpha \langl  r_E^2\rangl \over 3m_\pi}+\lim_{\bm q\to 0}{2\alpha \over \bm q^{\,2}} \int_{0}^\infty d t \bigg[Q_{11}^{inel}(\bm q,t) -Q_{11}^{inel}(\bm 0,t) \bigg].
\label{eq:beta}
\end{align}
\normalsize
Here $\alpha=1/137$ is the fine structure constant. 
The formulas are in discrete Euclidean spacetime but we keep the time axis continuous for notational convenience.
Zero-momentum Breit frame is employed in the formula to mimic low-energy Compton scattering, where the initial and final   pions are at rest and the photons have purely spacelike momentum.
The formulas have a similar structure in that they both have an elastic contribution in terms of the charge radius and pion mass, and an inelastic contribution in the form of subtracted time integrals. They differ in two aspects. The  $Q_{44}$ in $\alpha_E$ includes both elastic and inelastic contributions whereas the $Q^{inel}_{11}$ in $\beta_M$ includes only inelastic contributions.
In $\alpha_E$, the elastic $Q^{elas}_{44}(\bm q,t)$ is subtracted, whereas in $\beta_M$  it is the zero-momentum inelastic $Q_{11}^{inel}(\bm 0,t) $ that is subtracted.

Both $\alpha_E$ and $\beta_M$ have the expected physical unit of $a^3$ (fm$^3$).
 In the elastic term $\langl  r_E^2\rangl$ scales like $a^2$ and $m_\pi$ like $a^{-1}$.
In the inelastic term  $1/\bm q^{\,2}$ scales like $a^2$, $t$ scales like $a$,  and  $Q_{44}$ and $Q_{11}$ are dimensionless by definition. 
The $\alpha_E$  has been studied thoroughly in a previous work~\cite{Lee:2023rmz}, 
from which we take the results for pion mass $m_\pi$ and charge radius $ \langl  r_E^2\rangl $ and $\alpha_E$.
In this work we focus on the $\beta_M$ in Eq.~\eqref{eq:beta}.

The four-point function $Q_{11}$ is defined as,
\fontsize{9.5}{9.5}
\begin{align}
&Q_{11}(\bm q,t_3,t_2,t_1,t_0)  \equiv  \label{eq:Q11} \\&
 \frac{ \displaystyle\sum_{\bm x_3,\bm x_2,\bm x_1,\bm x_0} e^{-i\bm q\cdot \bm x_2} e^{i\bm q\cdot \bm x_1} 
\langl \Omega | \psi (x_3) :j^L_1(x_2) j^L_1(x_1):  \psi^\dagger (x_0) |\Omega \rangl }
{\displaystyle\sum_{\bm x_3,\bm x_0} \langl \Omega  | \psi (x_3) \psi^\dagger (x_0) |\Omega \rangl }. \nonumber
\end{align}
In this expression, $\Omega$ denotes the vacuum, and normal ordering is used to include the required subtraction of vacuum expectation values (VEV) on the lattice. 
The sums over $\bm x_0$ and $\bm x_3$ enforce zero-momentum pions at the source ($t_0$) and sink ($t_3$). 
The sum over $\bm x_1$ injects momentum $\bm q$ by the current at $t_1$, 
whereas sum over $\bm x_2$ takes out $\bm q$ by the current at $t_2$ to satisfy energy-momentum conservation in the process.
The two possibilities of time ordering are implied in the normal ordering. The time $t$ in Eq.\eqref{eq:alpha} and Eq.\eqref{eq:beta} represents the separation between the two currents $t=t_2-t_1$ with the two fixed ends $t_0$ and $t_3$ implied.

 We consider $\pi^+$  with standard  interpolating field,
  \normalsize
\beq
\psi_{\pi^+}(x)=\bar{d}(x) \gamma_5 u (x),
\label{eq:op}
\eeq
For the lattice version of electromagnetic current density in the $x$-direction,
we consider two options. One is a local current (or point current) built from up and down quark fields,
\beq
j^{(PC)}_1\equiv \,i\,Z_V \kappa\left(q_u \bar{u}\gamma_1 u + q_d \bar{d}\gamma_1 d \right).
\label{eq:j1PC}
\eeq
The factor $i$ here is needed to ensure that the spatial component $j^{(PC)}_1$ is hermitian, in contrast to the time component $j^{(PC)}_4$ in the electric case~\cite{Lee:2023rmz}. 
The reason is $(\bar{u}\gamma_1 u)^\dagger=- \bar{u}\gamma_1 u$  whereas  $(\bar{u}\gamma_4 u)^\dagger=\bar{u}\gamma_4 u$ (recall $\bar{u}\equiv u^\dagger \gamma_4$).
The factor $\kappa$ is to account for the quark-field rescaling $\psi\to \sqrt{2\kappa} \psi$ in Wilson fermions. The factor 2 is canceled by the 1/2 factor in the definition of the vector current ${1\over 2}\bar{\psi}\gamma_\mu \psi$.
The charge factors are $q_u=2/3$ and $q_d=-1/3$ where the resulting $e^2=\alpha$ in the four-point function has been absorbed in the definition of $\beta_M$ in  Eq.\eqref{eq:beta}. 
The advantage of this operator is that it leads to simple correlation functions. 
The drawback is that the renormalization constant for the vector current has to be determined. 
The other option is the conserved vector current for Wilson fermions on the lattice ($Z_V\equiv 1$) in  point-split form,
\beqs
 &j^{(PS)}_1 (x) \equiv \\&
\quad i\,q_u {\kappa_u} \big[ 
-\bar{u}(x) (1-\gamma_\mu) U_1(x) u(x+\hat{1}) \\& \qquad\qquad
+
\bar{u}(x+\hat{\mu}) (1+\gamma_1 ) U_1^\dagger(x) u(x) 
\big] 
\\ &
+  i\,q_d {\kappa_d} \big[ 
-\bar{d}(x) (1-\gamma_1) U_1(x) d(x+\hat{1}) \\&  \qquad \qquad
+ 
\bar{d}(x+\hat{1}) (1+\gamma_1 ) U_1^\dagger(x) d(x) 
\big].
\label{eq:j1PS}
\eeqs
Although conserved current explicitly involves gauge fields and lead to more complicated correlation functions, it has the advantage of circumventing the renormalization issue. All results in this work  are based on conserved current. 
\begin{figure}[b!]
\includegraphics[scale=0.4]{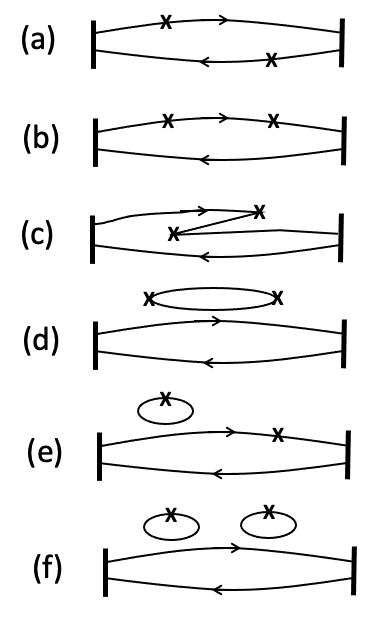}
\caption{Quark-line  diagrams of a four-point function contributing to polarizabilities of a charged pion.
Diagrams a,b,c are connected contributions whereas diagrams d,e,f are disconnected contributions. In each diagram, flavor permutations are assumed as well as gluon lines that connect the quark lines. The zero-momentum pion interpolating fields are represented by vertical bars (wall sources).}
\label{fig:4pt}
\end{figure}

At the quark level,  Wick contractions of quark-antiquark pairs in $Q_{11}$ in  Eq.\eqref{eq:Q11} lead to topologically 
distinct quark-line diagrams shown in Fig.~\ref{fig:4pt}.
We focus on the connected contributions in this study. 
The total connected contribution is simply the sum of the individual normalized terms,
\beqs
Q_{11}(\bm q,t_2,t_1)&=Q^{(a)}_{11}+ Q^{(b)}_{11}+Q^{(c)}_{11},
\label{eq:Q11abc}
\eeqs
for either point current or conserved current. The charge factors and flavor-equivalent contributions have been included in each diagram.
The disconnected contributions are more challenging and are left for future work.

\section{Simulation details and results}
\label{sec:results}
We use quenched Wilson action with $\beta=6.0$ and $\kappa=0.1520,\;0.1543,\; 0.1555,\; 0.1565$ on the lattice $24^3\times 48$.
We analyzed 500 configurations for $\kappa=0.1520$ and 1000 configurations each for rest of the kappas.
The scale of this action has been determined in Ref.~\cite{CABASINO1991195}, with inverse lattice spacing $1/a=2.312$ GeV and kappa critical $\kappa_c=0.15708$.
Dirichlet (or open) boundary condition is imposed in the time direction, while periodic boundary conditions are used in spatial dimensions.
The pion source is placed at $t_0=7$ and sink at $t_3=42$ (time is labeled from 1 to 48). One current is inserted at a fixed time $t_1$, while the other current $t_2$ is free to vary.
We consider four different combinations of momentum $\bm q=\{0,0,0\},\,\{0,0,1\},\,\{0,1,1\},\,\{0,0,2\}$. 
In lattice units they correspond to  the values ${\bm q}^2a^2=0,\, 0.068,\, 0.137,\,  0.274$, or in physical units to ${\bm q}^2=0,\, 0.366,\, 0.733,\, 1.465$ (GeV$^2$).

\subsection{Raw correlation functions}
\begin{figure*}[t!]
\includegraphics[scale=0.55]{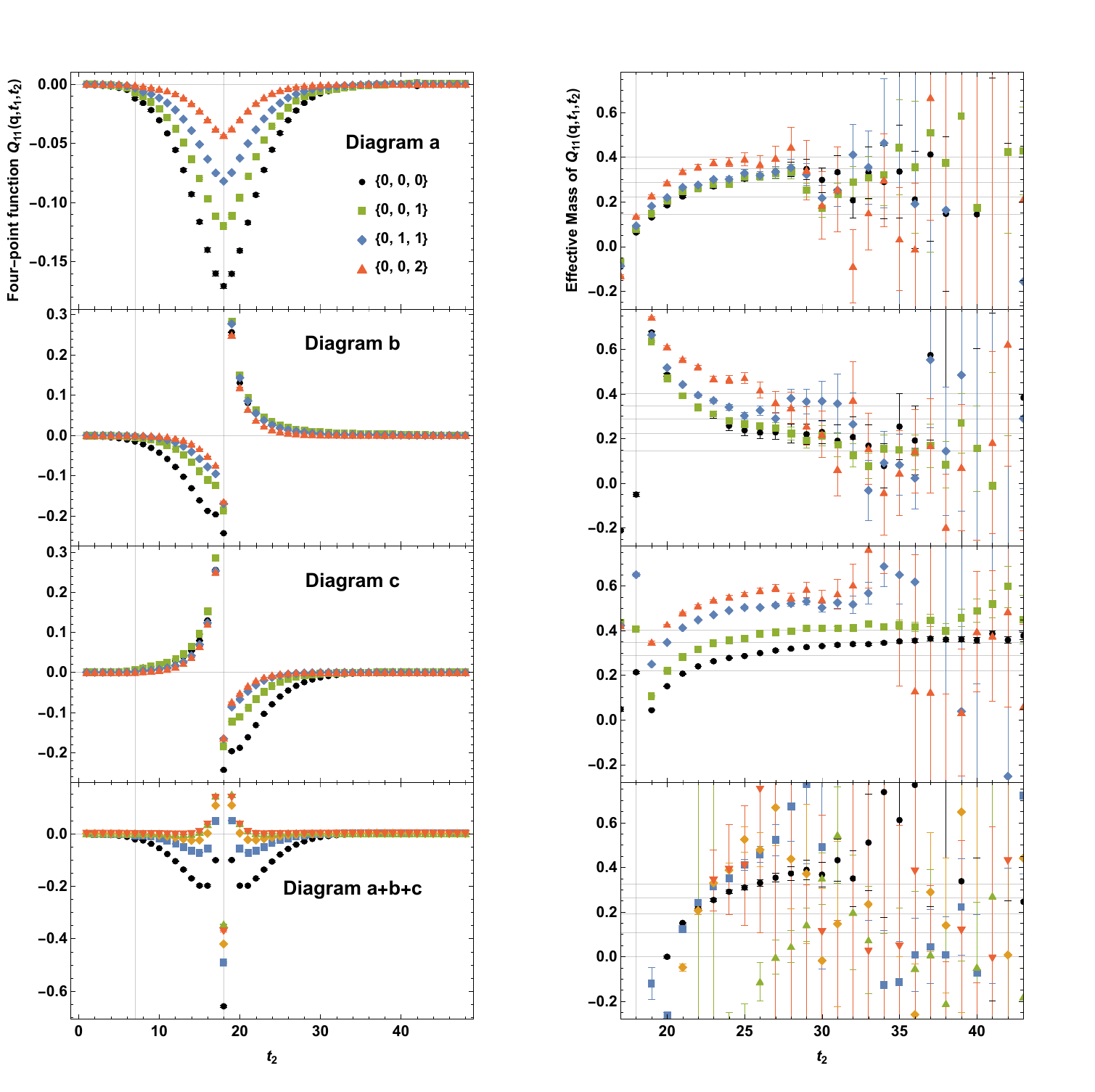}
\caption{
Individual and total  four-point functions (left ) and their effective mass functions (right) from the connected diagrams as a function of current separation at $m_\pi=600$ MeV. 
Vertical gridlines indicate the pion walls ($t_0=7$ and $t_3=42$) and the  fixed current insertion ($t_1=18$).
Horizontal gridlines in the effective mass functions are $E_\rho-m_\pi$ in lattice units where 
$E_\rho=\sqrt{\bm q^2+m_\rho^2}$ 
 with measured $m_\pi$ and $m_\rho$. 
The results in the total between $t_2=19$ and $t_2=41$ will be the signal for our analysis.
}
\label{fig:Q11PS}
\end{figure*}

In Fig.~\ref{fig:Q11PS} we show the raw normalized four-point functions, both individually and collectively,  at the four different values of momentum $\bm q$ and at $m_\pi=600$ MeV. 
All points are included and displayed on a linear scale for comparison purposes.
The special point of $t_1=t_2$ is regular in diagram a, but gives irregular results in diagram b and c at all values of $\bm q$.  The same irregularity is observed in the electric case. It is an unphysical contact interaction on the lattice which vanishes in the continuum limit. We treat this point with special care in our analysis below. The results about $t_1=18$ in diagram b and c are mirror images of each other,  simply due to the fact that they are from the two different  time orderings of the same diagram. In principle, this property could be exploited to reduce the cost of simulations by placing $t_1$ in the center of the lattice. In this study, however, we computed all three diagrams separately, and add them between $t_1=19$ and $t_3=41$ as the signal.

To see the structure of the four-point function in Eq.\eqref{eq:Q11}, we insert a complete set of intermediate states in the numerator (trice) and in the denominator (once), and make use of translation invariance and kinematics,
%
\fontsize{8.5}{8.5}
\beqs
& Q_{11}(\bm q,t_3,t_2,t_1,t_0)  =  \sum_{\bm x_3,\bm x_2,\bm x_1,\bm x_0}  N^3_s \sum_{n,n_i,n_f} \\&
\langl \Omega |\psi(\bm 0) | n_f(\bm 0) \rangl  e^{-am_\pi (t_3-t_2)}  \\&
\langl n_f(\bm 0) | j^L_1(0) | n(\bm q) \rangl  e^{-aE_n(\bm q) (t_2-t_1)} \\&
\langl n(\bm q) | j^L_1(0) | n_i(\bm 0) \rangl e^{-am_\pi (t_1-t_0)} \langl n_i(\bm 0) | \psi^\dagger(\bm 0)|\Omega \rangl  \\&
/  \big[ \sum_{\bm x_3,\bm x_0} N_s \sum_{n} 
\langl \Omega |\psi(\bm 0) | n(\bm 0) \rangl e^{-am_n (t_3-t_0)}   \langl  n(\bm 0) |\psi^\dagger(\bm 0) | \Omega \rangl \big]  \\ &
-  N_s \sum_{n} 
\langl \Omega | j^L_1(0) | n(\bm q) \rangl  e^{-aE_n(\bm q) (t_2-t_1)} \langl n(\bm q) | j^L_1(0)  | \Omega \rangl  \\ & 
 =  -N^2_s 
| \langl \pi(\bm 0) | j^L_1(0) | \pi(\bm q) \rangl|^2  e^{-a(E_\pi-m_\pi) (t_2-t_1)}  \\  &
\quad +N_s |\langl \Omega | j^L_1(0) | \pi(\bm q) \rangl |^2  e^{-aE_\pi (t_2-t_1)} + \cdots,
\label{eq:Q1} 
\eeqs
%
where the leading contributions are isolated in the last step under time limits $t_3\gg t_{1,2} \gg t_0$.
The change of sign is due to the metric factor for spatial components in Eq.\eqref{eq:j1PC}.
The $N_s=N_xN_yN_z$ is the number of spatial sites on the lattice. 
The normal ordering in Eq.\eqref{eq:Q11} is formally defined as,
\beqs
&\langl \Omega | \psi (x_3) :j^L_1(x_2) j^L_1(x_1):  \psi^\dagger (x_0) |\Omega \rangl  \\&
\equiv \langl \Omega | \psi (x_3) T[j^L_1(x_2) j^L_1(x_1)]  \psi^\dagger (x_0) |\Omega \rangl  \\&
 -  \langl \Omega | T j^{L}_1(x_2) j^{L}_1(x_1) | \Omega\rangl \langl \Omega | \psi (x_3)  |  \psi^\dagger (x_0)|\Omega \rangl,
\eeqs
where $T$ in the first term means time-ordering and the second term signifies a subtraction of vacuum expectation values (VEV) on the lattice in the disconnected diagrams.
In fact, subtraction is only needed for diagram d; it vanishes for diagrams e and f.
The pion two-point function cancels exactly in the second term in Eq.\eqref{eq:Q1}.

Poin form factor is contained in the elastic matrix element, 
\beqs
&\langl\pi(p')| j^L_\mu(0) | \pi(p)\rangl
\\&
=(p'+p)_\mu F_\pi(q^2) +q_\mu {p'^2-p^2\over q^2}(1-F_\pi(q^2)).
\eeqs
It vanishes for $j^L_1$  as long as  $(p'+p)_\mu$ does not have a $\mu=1$ component. The condition is indeed satisfied under the zero-momentum Breit frame and our selection of  $\bm q$ values. This is the reason that $\bm q=\{1,1,1\}$ is excluded from the set of $\bm q$ values  relative to the electric case.

In other words, there is no elastic contribution in the second term of $\beta_M$  in Eq.\eqref{eq:beta} as long as transverse momentum to $j^L_1$ is considered. This is evident in the effective mass functions in Fig.~\ref{fig:Q11PS} where the intermediate states are not on-shell pions, but states with different mass and energy.
Possible intermediate states are either vector or axial mesons in the magnetic channel. For reference, we draw horizontal lines
$E_\rho-m_\pi$ in lattice units where 
$E_\rho=\sqrt{\bm q^2+m_\rho^2}$, using  
 measured $m_\pi$ and $m_\rho$. 
 The effective mass functions in Fig.~\ref{fig:Q11PS} are only provided for reference purposes on the intermediate state.  They can become noisy at large current separations and higher momentum. This is not a concern since there is no fitting performed at large times. The signal is the time integral of subtracted four-point functions, which amounts to evaluating the area between two curves. And the signal is dominant at small times.

\subsection{Magnetic polarizability}
\begin{figure}[h!]
\includegraphics[scale=0.45]{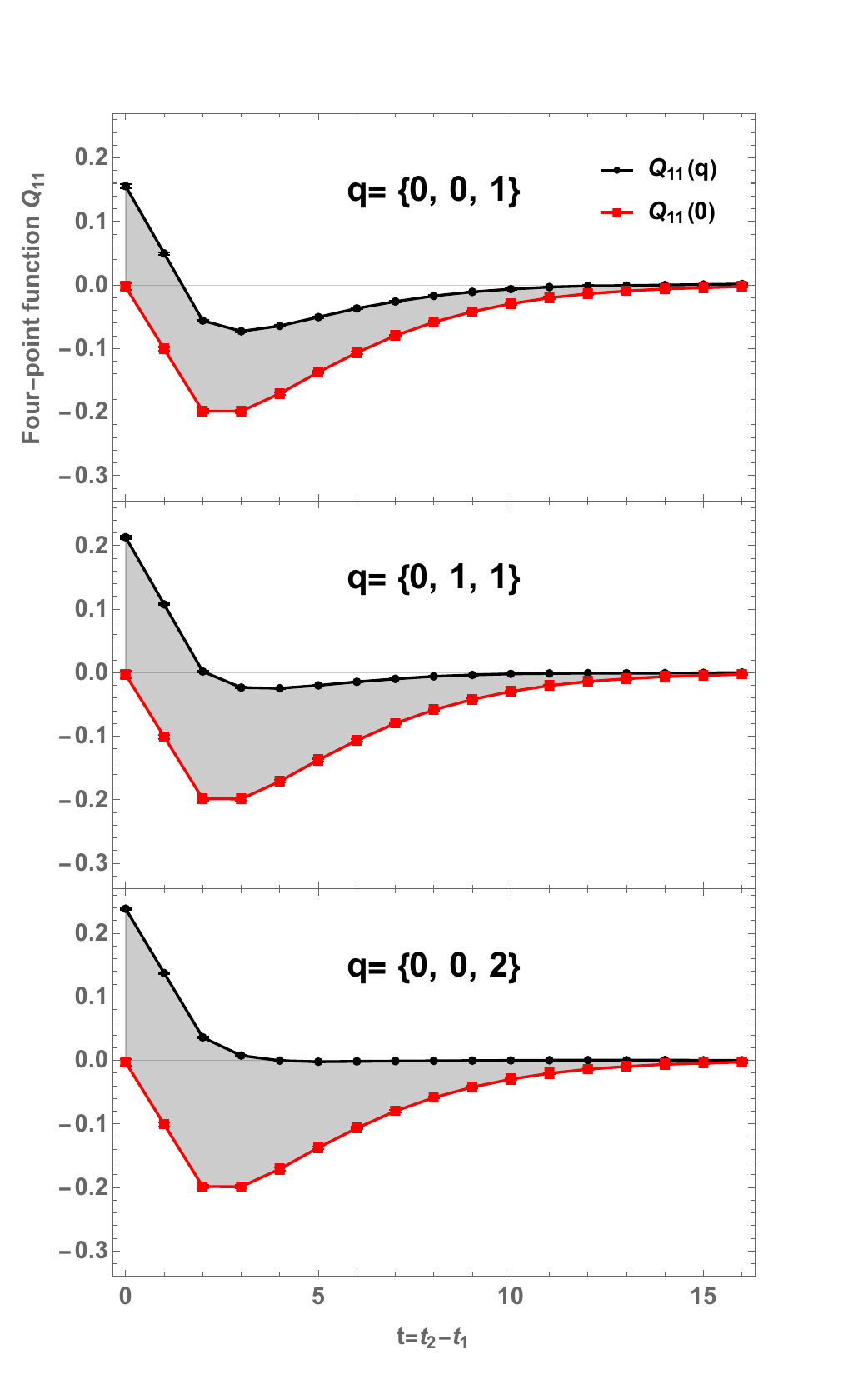}
\caption{
Momentum-carrying $Q_{11}(\bm q)$ and zero-momentum  $Q_{11}(\bm 0)$  at different values of $\bm q$ at $m_\pi=600$ MeV.
 The shaded area between the two is the dimensionless signal contributing to magnetic polarizability.
}
\label{fig:QQ}
\end{figure}
\begin{figure}[b!]
\includegraphics[scale=0.45]{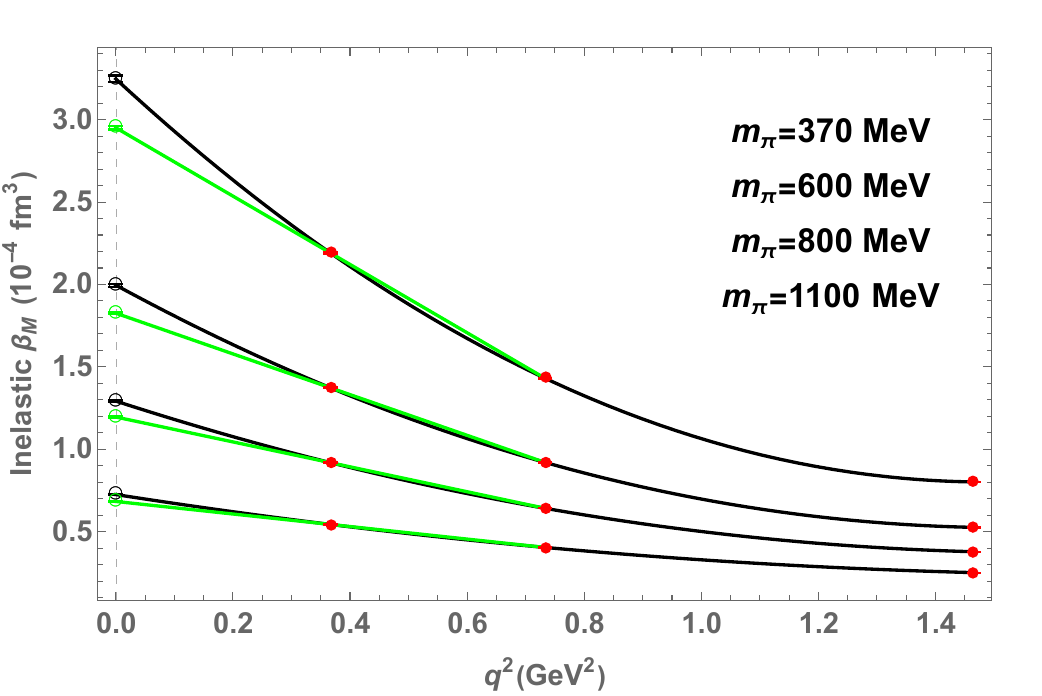}
\caption{
Momentum dependence of the inelastic term in Eq.~\eqref{eq:beta} and its extrapolation to $\bm q^2=0$ at all pion masses.
Red points are based on the shaded areas in Fig.~\ref{fig:QQ}. Black curve is a quadratic extrapolation using all three points. Green curve is a linear extrapolation based on the two lowest points. Empty points indicate the corresponding extrapolated values contributing to $\beta_M$.
}
\label{fig:Qzero}
\end{figure}

In Fig.~\ref{fig:QQ} we show the connected contribution $Q_{11}(\bm q)$ at different $\bm q$ values and  zero-momentum $Q_{11}(\bm 0)$ as a function of current separation $t=t_2-t_1$ in lattice units.  Only results for $m_\pi=600$ MeV are shown as an example; the graphs at the other pion masses look similar.  The time integral required for $\beta_M$ in the formula,
  $(1/a)\int dt \big[ Q_{11}(\bm q,t)-Q_{11}(\bm 0,t)\big]$,  is simply the shaded area between the two curves, and it is positive.  
One detail to notice is that the curves include the $t=0$ point which has unphysical contributions in $Q_{11}$ mentioned earlier. We would normally avoid this point and only start the integral from $t=1$. However, the chunk of area between $t=0$ and $t=1$ is the largest piece in the integral. To include this contribution, we linearly extrapolated both  $Q_{11}(\bm q)$  and  $Q_{11}(\bm 0)$ back to $t=0$ using the two points at $t=1$ and $t=2$. As the continuum limit is approached, the $t=0$ point will become regular and the chunk will shrink to zero.


%
\begin{figure}[tbh!]
\includegraphics[scale=0.45]{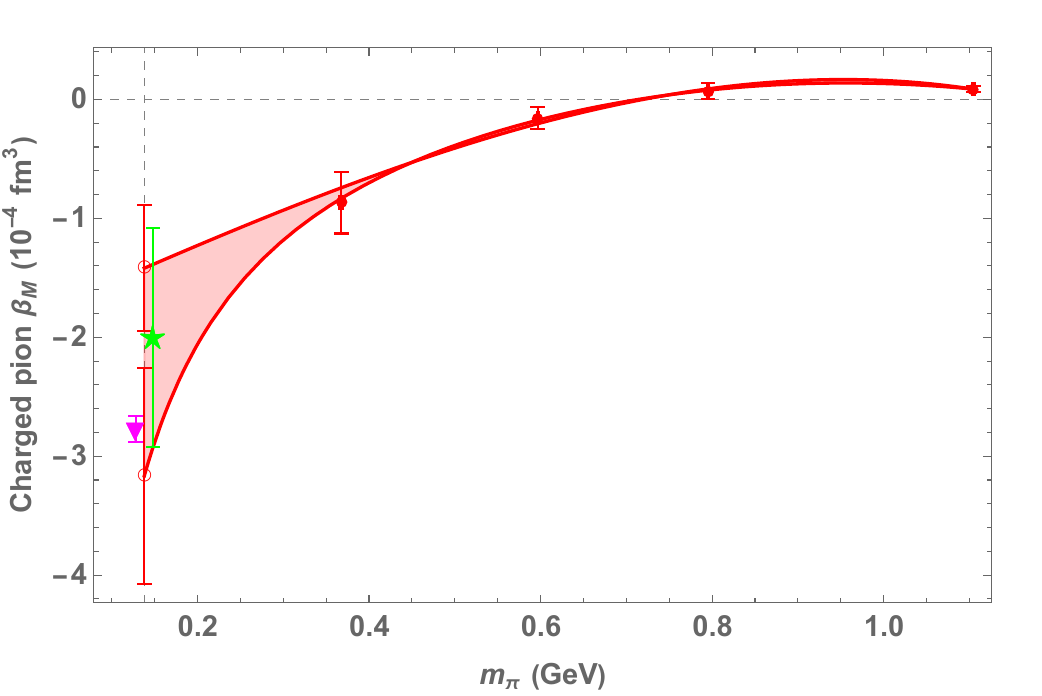}
\caption{
Chiral extrapolation of charged pion magnetic polarizability. 
 For better viewing, the PDG value (star) and ChPT value (triangle) are shifted horizontally by 10 MeV.
}
\label{fig:chiral}
\end{figure}
\begin{figure}[tbh!]
\includegraphics[scale=0.5]{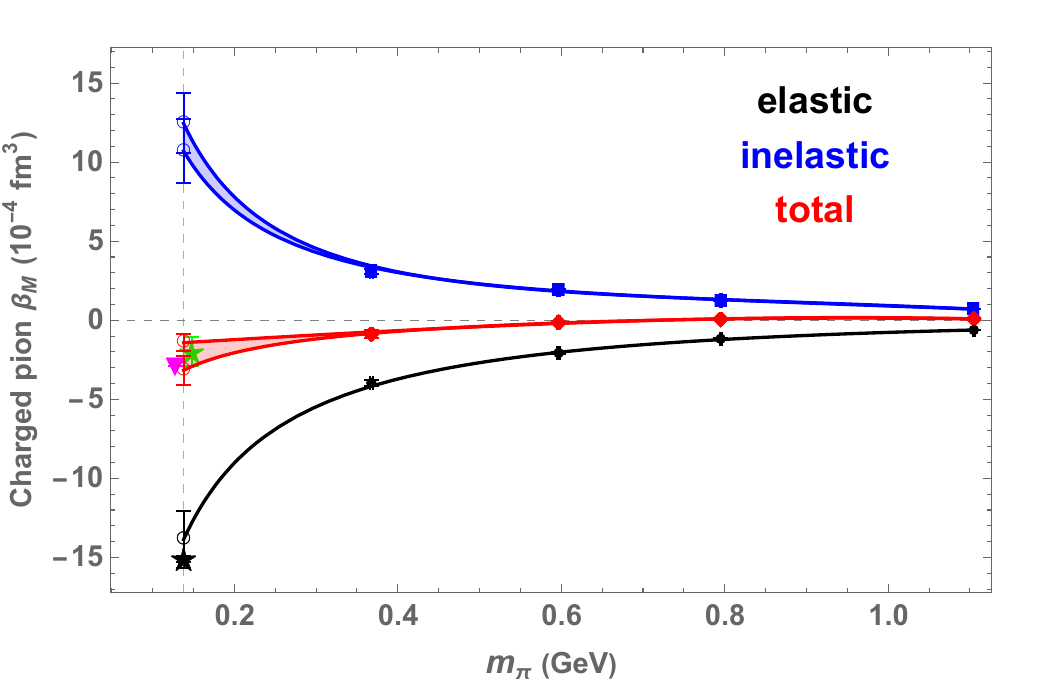}
\caption{
Individual and total contributions to charged pion $\beta_M$ from  four-point functions in lattice QCD based on the formula in  Eq.\eqref{eq:beta}.
The total is taken from Fig.~\ref{fig:chiral}, and the elastic from Ref.~\cite{Lee:2023rmz}; both are chirally extrapolated to the physical point.
The inelastic is from the difference of the two.
}
\label{fig:beta}
\end{figure}
The inelastic term can now be constructed by multiplying ${2\alpha / \bm q^{\,2}}$ and the time integral,  and it is a function of momentum. Since $\beta_M$ is a static property, we  extrapolate it to $\bm q^2=0$ smoothly. We consider two fits, a quadratic fit $a+b\, x+c\, x^2$  ($x=\bm q^2$) using all three  data points, and a linear fit using the two lowest points. The results are shown in Fig.~\ref{fig:Qzero} for all pion masses. One observes a spread in the extrapolated values at $\bm q^2=0$. We treat the spread as a systematic effect as follows.  We take the average of the two  extrapolated values along with statistical uncertainties, and  half of the difference in their central values as a systematic uncertainty. The statistical and  systematic uncertainties are then propagated in quadrature  to the analysis of  $\beta_M$.
For our data, the statistical uncertainties are relatively small, so the systematic uncertainties are dominant in the inelastic contribution.

\begin{table*}
\caption{Summary of results in physical units from two-point and four-point functions. 
Results for charge radius and $\alpha_E$ are taken from previous work~\cite{Lee:2023rmz}. 
Elastic $\beta_M$ and total $\beta_M$ are chirally extrapolated to the physical point. Inelastic $\beta_M$ at the physical point is taken as the difference of the two.
Known values  from ChPT and PDG are listed for reference.
All polarizabilities are in units of $10^{-4}\;\text{fm}^3$.
}
\label{tab:final}
\begin{tabular}{c}
$      
\renewcommand{\arraystretch}{1.2}
\fontsize{8}{8}
\begin{array}{l|ccccc|c}
\hline
  & \text{$\kappa $=0.1520} & \text{$\kappa $=0.1543} & \text{$\kappa $=0.1555} & \text{$\kappa $=0.1565} & \text{physical point} & \text{known value}
   \\
\hline
 m_{\pi }\text{ (MeV)} & 1104.7\pm 1.2 & 795.0\pm 1.1 & 596.8\pm 1.4 & 367.7\pm 2.2 &  138 & 138 \\
 m_{\rho }\text{ (MeV)} & 1273.1\pm 2.5 & 1047.3\pm 3.4 & 930.\pm 7. & 830.\pm 17. & 770 & 770\\
 \hline
 \langl r_E^2\text{$\rangl$ (}\text{fm}^2\text{) } & 0.1424\pm 0.0029 & 0.195\pm 0.007 & 0.257\pm 0.005 & 0.304\pm 0.016  & 0.40\pm 0.05 & 0.434\pm 0.005 \text{ (PDG)}\\
     \hline
 \alpha _E\text{ elastic}  & 0.618\pm 0.012 & 1.17\pm 0.04 & 2.07\pm 0.04 & 3.97\pm 0.21 & 13.9\pm 1.8 & 15.08\pm 0.13 \text{ (PDG)}\\
  \alpha _E\text{ inelastic }  & -0.299\pm 0.019 & -0.672\pm 0.030 & -0.92\pm 0.11 & -1.27\pm 0.13 & -9.7\pm 1.9 \text{ to} -5.1\pm 2.0 &\\
 \alpha _E\text{ total } & 0.319\pm 0.023 & 0.50\pm 0.05 & 1.15\pm 0.11 & 2.70\pm 0.25 & 4.2\pm  0.5 \text{ to } 8.8\pm 0.9 & 2.93\pm 0.05 \text{ (ChPT)}\\
      &   &   &   &  & & 2.0\pm 0.6 \pm 0.7\text{ (PDG)}\\
        \hline
    \beta _M\text{ elastic} & -0.618\pm 0.012 & -1.17\pm 0.04 & -2.07\pm 0.04 & -3.97\pm 0.21 & -13.9\pm 1.8 & -15.08\pm 0.13 \text{ (PDG)}\\
 \beta _M\text{ inelastic } & 0.705\pm 0.021 & 1.24\pm 0.05 & 1.91\pm 0.09 & 3.10\pm 0.15 & 10.7\pm 2.0 \text{ to } 12.4\pm 1.9&\\
 \beta _M\text{ total } & 0.087\pm 0.024 & 0.07\pm 0.06 & -0.16\pm 0.09 & -0.87\pm 0.26 & -3.2\pm 0.9 \text{ to} -1.4\pm 0.5 & -2.77\pm 0.11 \text{ (ChPT)}\\
       &   &   &   &  & & -2.0\pm 0.6 \pm 0.7\text{ (PDG)}\\
  \hline
\end{array}
$   
\end{tabular}
\end{table*}
\normalsize

Finally, we assemble the two terms in the formula in Eq.\eqref{eq:beta} to obtain $\beta_M$ in physical units.
The results are summarized  in Fig.~\ref{fig:chiral}  and in Table~\ref{tab:final}. 
At each pion mass the elastic term is  negative, whereas the inelastic term is positive.  The total is slightly positive at the two heaviest pion masses, then turns negative as the pion mass is lowered.  
To see how the trend continues to smaller pion masses, we take the total values for $\beta_M$ at the four pion masses and perform a smooth extrapolation to the physical point.  Since our pion passes are relatively large, we consider two forms to cover the range of uncertainties in the extrapolation: 
 a polynomial form $a+b\, m_\pi+c\, m_\pi^3$ and a form with a divergent $1/m_\pi$ term ${a\over m_\pi}+b\, m_\pi+c\, m_\pi^3$  inspired by ChPT~\cite{Gasser_2006,BURGI1996392}. 
The spread can be considered as a systematic effect.
The extrapolated value of $-3.2\pm 0.9 \text{ to} -1.4\pm 0.5$ at the physical point is comparable to  the known value of  $-2.0\pm 0.6\pm 0.7$ from PDG~\cite{Workman:2022ynf}  and  $-2.77(11)$ from two-loop contribution of ChPT~\cite{Gasser_2006,Moinester:2019sew}.  
An interesting feature is a sign change from positive to negative as pion mass is lowered. It happens around 750 MeV. In contrast, there is no sign change in the electric case.

To get an overview on how the $\beta_M$ comes about, we show in Fig.~\ref{fig:beta} three terms on the same graph: elastic, inelastic, and their sum.  
The inelastic curve is taken as the difference of the total and the elastic curves. This gives a constraint of  $10.7\pm 2.0 \text{ to } 12.4\pm 1.9$  for the inelastic at the physical point. It would be interesting to verify this chiral behavior in the inelastic term directly in future simulations.
The results in this figure  point to the following physical picture: $\beta_M$ is the result of a large cancellation between the elastic and inelastic contributions. The cancellation is more significant  than in the electric case~\cite{Lee:2023rmz}. 
This cancellation appears to continue in the approach to the physical point, resulting in a total value that is relatively small and negative, and  a relatively mild pion mass dependence compared to the individual contributions.  It is almost the complete opposite to the electric case~\cite{Lee:2023rmz}. 

 A comparison on $\beta_M$ can be made here between the four-point function method and  the background field method.
 For the former, our value of $-3.2(9)$ is the only attempt at the moment. For the latter, there are several calculations.
 In Ref.~\cite{Luschevskaya_2015,Luschevskaya:2015cko},  $\beta_M$ is studied for both charged and neutral pions.
 A fitting form is used that includes Landau levels and up to $B^4$ contributions in magnetic field for charged pions. Values of $-1.15(31)$ and $-2.06(76)$ are obtained on two different lattices. No chiral extrapolation is performed. Since only bare quark masses are given we could not ascertain what pion masses they correspond to. 
In Ref.~\cite{He:2021eha},  a Laplacian-mode projection technique is employed at the quark propagator level to filter out the Landau levels.  
The same technique is used on the nucleon~\cite{Bignell:2020xkf}.
A final value of $-1.70(14)(25)$ is reported. It also predicts a sign change in $\beta_M$, but only after chiral extrapolation. The simulated results are positive at all the pion masses considered, down to about 300 MeV.  A Pad\'e form is introduced to extrapolate the positive values to the negative one at the physical point. The sign change happens at around 225 MeV.
This is different from the sign change observed in Fig.~\ref{fig:beta}, which happens at a heavier pion mass, before chiral extrapolation.  
This is an interesting  puzzle for future investigations. The resolution could be in the different systematics present in the two calculations. For the four-point function method in this work, it could be the quenched approximation, disconnected diagrams, and the contact term in the connected diagrams.

\subsection{  $\alpha_E+\beta_M$ }
\begin{figure}[ht]
\includegraphics[scale=0.45]{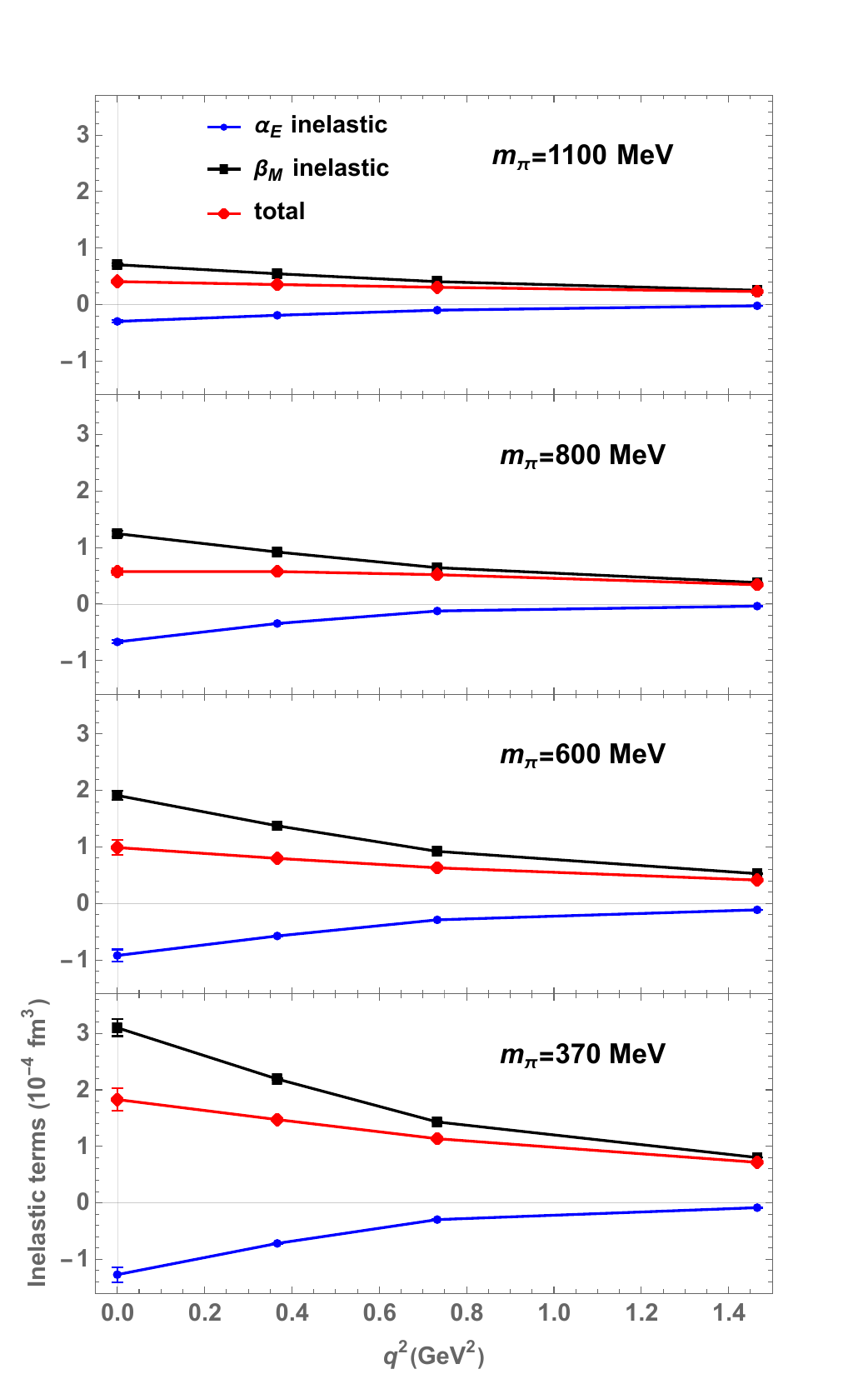}
\caption{
Momentum dependence of the inelastic terms at different pion masses. 
The values at $\bm q^2=0$ are from a linear extrapolation using the two lowest points. The curves are straight lines connected all the points. 
The sum of the two inelastic terms (red)  is a direct measure of  the momentum dependence  for $\alpha_E(\bm q)+\beta_M(\bm q)$.
}
\label{fig:QzeroAll}
\end{figure}
\begin{figure}    
\includegraphics[scale=0.5]{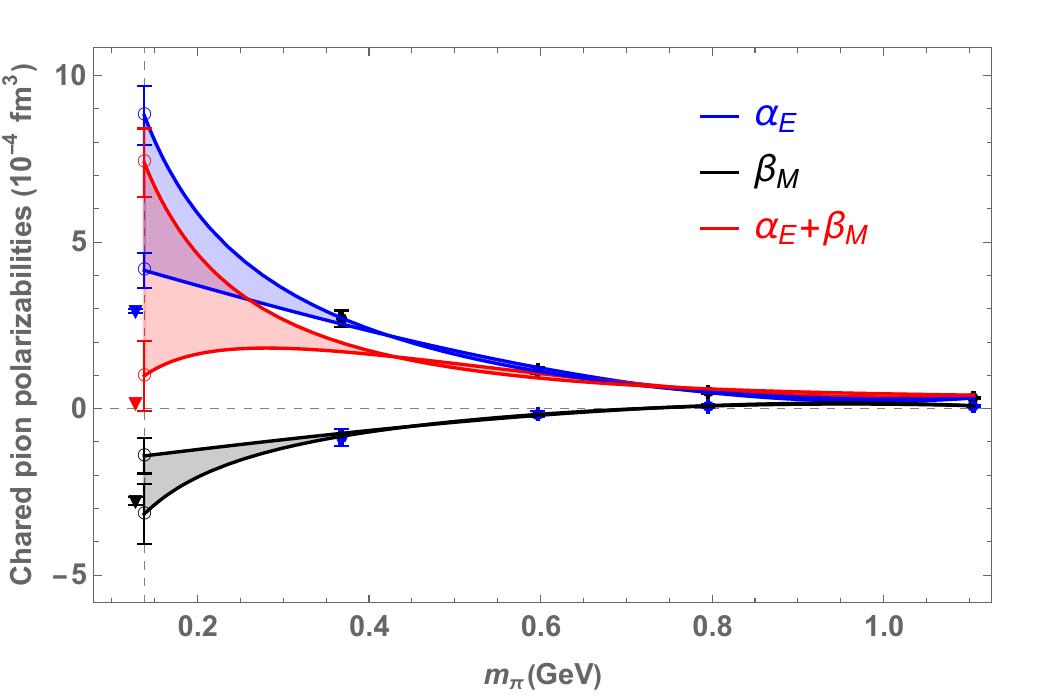}
\caption{
Pion mass dependence of individual and sum of polarizabilities  for  a charged pion from  four-point functions in lattice QCD.
The $\alpha_E$ is previously determined from the formula in Eq.\eqref{eq:alpha} and $\beta_M$ in the present study from Eq.\eqref{eq:beta}. 
The curves are nonlinear chiral extrapolations. The color-coded triangles  are values from  ChPT and they are shifted horizontally by 10 MeV for better viewing.}
\label{fig:ab}
\end{figure}

Here we take a closer look at the sum of electric and magnetic polarizabilities.
ChPT gives a solid  prediction that $\alpha_E+\beta_M\approx 0$ at leading-oder and $\alpha_E+\beta_M\approx 0.16$ at the two-loop order~\cite{Gasser_2006} in units of $10^{-4} \text{ fm}^3$.  Baldin sum rule~\cite{BALDIN1960310,Moinester:2019sew} applied to a charged pion gives $\alpha_E+\beta_M\approx 0.39(4)$ in the same units.

In the four-point function formalism,   we note first that if we add Eq.\eqref{eq:alpha} and Eq.\eqref{eq:beta}, the elastic charge radius terms cancel exactly, leaving only inelastic contributions in the form of subtracted time integrals,
\fontsize{9}{9}
\beqs
&\alpha_E+\beta_M=
\lim_{\bm q\to 0}{2\alpha \over \bm q^{\,2}} \int_{0}^\infty d t\\
& \bigg[Q_{44}(\bm q,t) -Q^{elas}_{44}(\bm q,t) 
+Q_{11}^{inel}(\bm q,t) -Q_{11}^{inel}(\bm 0,t) \bigg].
\label{eq:ab}
\eeqs
\normalsize
This can be regarded as a sum rule for $\alpha_E+\beta_M$ on the lattice (instead of energy integration over cross sections, it is a time integration over subtracted four-point functions).
Second, the inelastic terms are opposite in sign so there is a cancellation in the inelastic contributions as well. Specifically, $\alpha_E$ inelastic is negative whereas $\beta_M$ inelastic is positive and both  are momentum-dependent.
In Fig.~\ref{fig:QzeroAll}, we show the momentum dependence  separately along with their sum including the extrapolated  $\bm q^2=0$ limit.
They are displayed on the same scale at different pion masses to facilitate comparison.
The magnetic points are taken from Fig.~\ref{fig:Qzero}. The electric points are taken from Ref.~\cite{Lee:2023rmz}. 
Note that we leave out the $\bm q=\{1,1,1\}$ point from the electric case for a one-to-one comparison.
The salient feature is that they are not only opposite in sign, but  the magnetic is consistently larger than the electric in magnitude, over  the entire momentum range.  As a result, the cancellation leaves a relatively small and positive value. The value at $\bm q^2=0$ appears to grow with decreasing pion mass, which is deviating from ChPT expectations.

Finally, we look at pion mass dependence, by adding inelastic terms at the $\bm q^2=0$ limit and elastic terms. We plot in Fig.\ref{fig:ab}  $\beta_M$ from this work, $\alpha_E$ from the previous work~\cite{Lee:2023rmz} , and their sum on the same graph. We see $\alpha_E$ is positive in the pion mass range studied, and $\beta_M$ is negative except at the heaviest pion mass of 1100 MeV. The  cancellation leads to a positive value for $\alpha_E$  $\alpha_E+\beta_M$ at the three lowest pion masses. 
It is unclear how the cancellation plays out in the approach to the physical point due to large uncertainties in the extrapolations.

\section{Conclusion}
\label{sec:con}

Building on the study of electric polarizability $\alpha_E$ for a charged pion using four-point functions in lattice QCD~\cite{Lee:2023rmz},
we investigated its magnetic polarizability $\beta_M$ using the same methodology and simulation parameters. 
The extension is relatively straightforward, mainly replacing charge-charge correlation ($Q_{44}$) 
with current-current correlation ($Q_{11}$). The formula for $\beta_M$ in Eq.\eqref{eq:beta}  has a similar structure to the one for $\alpha_E$ in 
Eq.\eqref{eq:alpha}. They share the same charge radius $\langl r_E^2\rangl$ and pion mass in  the elastic contribution, but this term appears with an opposite sign in the two formulas.   
The inelastic contribution is in the form of a subtracted time integral. In the electric case, it is the elastic $Q_{44}^{elas}(\bm q)$ at each momentum that is subtracted from the total, whereas  in the magnetic case it is the zero-momentum inelastic  $Q_{11^{inel}}(\bm 0)$  that is subtracted. 
Only $Q_{44}$ is needed for $\alpha_E$, but both $Q_{44}$ and $Q_{11}$ are needed  for $\beta_M$  due to coupling between the two formulas.
The methodology requires two- and four-point functions,  but not three-point functions. 


The emerging picture in Fig.\eqref{fig:beta} for $\beta_M$ is similar to that for $\alpha_E$, but in the reverse sense: it is the result of a cancellation between a negative elastic contribution and a positive inelastic contribution. The cancellation is more significant than in the electric case.  Individually, each contribution has strong pion mass dependence in the approach to the chiral limit, but the total has a small { negative} value with a relatively mild pion mass dependence. 
Combining the results of this study and those of Ref.~\cite{Lee:2023rmz}, we found that $\alpha_E+\beta_M$ is the consequence of cancellations  at three levels to varying degrees.
First, there is an exact cancellation in the elastic terms. Second, there is a cancellation in the inelastic terms as a function of momentum, with the magnetic slightly larger than the electric, leaving a relatively weak momentum dependence at fixed pion mass. Third, at the static limit there is a partial cancellation between $\alpha_E$ and $\beta_M$ at the lowest three pion masses explored, leaving a positive value. Although the resulting sign of $\alpha_E+\beta_M$ is consistent with ChPT, it is unclear quantitatively in the approach to the physical point since chiral extrapolation of the results suffers from large uncertainties. These issues point to the importance of pushing to smaller pion masses.

We caution that the above picture is still subject to a number of systematic effects at the proof-of-principle stage, such as the quenched approximation, finite-volume effects, and disconnected loops.  In particular, there is a systematic effect in the connected diagrams from the contact term when the two currents overlap on the same quark. This is a lattice artifact of unknown size that can only be computed correctly very close to the continuum limit. 
Additionally, there is a puzzling difference between the  four-point function method  and the background field method that warrants further study.
Although both methods yield similar negative values at the physical point after chiral extrapolation, they have different signs at non-physical pion masses.
The resolution should focus on the different  systematic effects present in the two studies.

\vspace*{5mm}
\begin{acknowledgments}
This work was supported in part by U.S. Department of Energy under  Grant~No.~DE-FG02-95ER40907 (FL, AA) and UK Research and Innovation grant {MR/S015418/1} (CC). WW would like to acknowledge support from the Baylor College of Arts and Sciences SRA program.
AA would like to acknowledge support from University of Maryland. 
The calculations are carried out at DOE-sponsored NERSC at Livermore and NSF-sponsored TACC at Austin.
\end{acknowledgments}

\clearpage
\bibliography{xmag}

\begin{widetext}
\appendix

\section{Magnetic correlation functions}
\label{sec:cfun}

A detailed formalism and notation has been laid out in the study of electric polarizability~\cite{Lee:2023rmz}. 
Here we present essential equations needed for magnetic polarizability and point out subtle differences.

Eq.\eqref{eq:Q11} is a normalized four-point function where the normalization constant is taken as the wall-to-wall two-point function,
\fontsize{9}{9}
\beqs
&  \sum_{\bm x_3,\bm x_0} \langl \Omega | \psi (x_3) \psi^\dagger (x_0) |\Omega\rangl 
   =
\Tr_{s,c} \Bigg[ 
\bigg( {\cal W}^T P(t_3) V_{a1}^{(d)} \bigg)^\dagger 
\bigg({\cal W}^T P(t_3)  V_{a1}^{(u)}  \bigg) \Bigg]  
 =
\Tr_{s,c} \Bigg[ 
\bigg( {\cal W}^T P(t_0) V_{a2}^{(u)} \bigg)^\dagger 
\bigg({\cal W}^T P(t_0)  V_{a2}^{(d)}  \bigg) \Bigg].
\label{eq:2pt3}
\eeqs
Here $V_{a1}$ and  $V_{a2}$  are zero-momentum quark propagators emanating from the walls at $t_0$ and $t_3$, respectively,
\beq
V_{a1}^{(q)} \equiv  M_q^{-1} P(t_0)^T {\cal W}, \quad
V_{a2}^{(q)} \equiv  M_q^{-1} P(t_3)^T {\cal W}.
\label{eq:a1a2}
\eeq
Here $M_q^{-1}$ is the inverse quark matrix, $P(t)$ a projector that projects a quark propagator from a given source to  time slice $t$,  and ${\cal W}$ the wall source.

For diagram a,  the unnormalized four-point function with local  current (denoted as PC) is written as,
\beq
\tilde{Q}^{(a, PC)}_{11}(\bm q,t_1,t_2)
=-{4\over 9}Z_V^2\kappa^2 \Tr_{s,c} \Bigg[ \bigg(
 \left[P(t_2)V_{a2}\right]^\dagger \gamma_5 \gamma_{1} e^{i\bm q}  P(t_2) V_{a1}  
\bigg)^\dagger \bigg(
 \left[P(t_1)V_{a2}\right]^\dagger \gamma_5 \gamma_{1} e^{i\bm q}  P(t_1) V_{a1}  
\bigg) \Bigg].
\eeq
Comparing to $Q_{44}$ in the electric case~\cite{Lee:2023rmz},  in addition to replacing $\gamma_4$ with $\gamma_1$, there is an overall sign change  in $Q_{11}$ due to the $i$ factor in $j_1^L$ in Eq.\eqref{eq:j1PC}. 
For conserved current (denoted as PS),
\fontsize{9}{9}
\beqs
& \tilde{Q}^{(a,PS)}_{11}(\bm q,t_1,t_2)
=-{4\over 9} \kappa^2 \Tr_{s,c} \Bigg[ \\ & \bigg(
 \left[P(t_2)V_{a2}\right]^\dagger \gamma_5(1-\gamma_{1})e^{i\bm q}  U_{1}(t_2,t_2)  P(t_2) V_{a1}
-\left[P(t_2)V_{a2}\right]^\dagger \gamma_5(1+\gamma_{1}) U^\dagger_{1}(t_2,t_2) e^{i\bm q}  P(t_2) V_{a1}
\bigg)^\dagger \\ 
& \bigg(
 \left[P(t_1)V_{a2}\right]^\dagger \gamma_5(1-\gamma_{1})e^{i\bm q}  U_{1}(t_1,t_1)  P(t_1) V_{a1}
-\left[P(t_1)V_{a2}\right]^\dagger \gamma_5(1+\gamma_{1}) U^\dagger_{1}(t_1,t_1) e^{i\bm q} P(t_1) V_{a1}
\bigg) \Bigg].
\eeqs
In our notation, the current split in space is only implicitly carried in the gauge links, not in quark propagators, whereas the split in time is explicitly carried in both the propagators and gauge links. For example, a split at time slice $t_2$ has the following  meaning in the links,
 \beqs
 U_{\mu}(t_2,t_2+\hat{\mu}_4) &\equiv
 \begin{cases} U_4(t_2,t_2+1),  & \text{if } \mu= 4\\ 
 U_\mu(t_2,t_2), &\text{if } \mu \neq 4,
 \end{cases} \\ 
  U_{\mu}^{\dagger}(t_2+\hat{\mu}_4,t_2) &\equiv
 \begin{cases} U_4^{\dagger}(t_2+1,t_2), & \text{if } \mu= 4\\ 
 U_\mu^{\dagger}(t_2,t_2), &\text{if } \mu \neq 4.
 \end{cases}
 \eeqs
Consequently, $U_1$ and $U_1^\dagger$ do not commute with $e^{i\bm q}$, unlike $U_4$ and $U_4^\dagger$ in the electric case. 

For diagram b and local current,  we have 
\normalsize
\beqs
\begin{aligned}
\tilde{Q}^{(b, PC)}_{11}(\bm q,t_2)
&={5\over 9}Z_V^2\kappa^2  \Tr_{s,c} \Bigg[ 
\left[P(t_2)\gamma_5 V_{a3}^{(1,PC)}(\bm q)\right]^\dagger  \gamma_1 e^{-i\bm q} P(t_2) V_{a2} {\cal W}^T P(t_3) \gamma_5  V_{a1} \Bigg], \\
\end{aligned} 
\eeqs
where $V_{a3}^{(1,PC)}$ is a SST quark propagator built from $V_{a1}$, 
\beq 
V_{a3}^{(1,PC)}(\bm q) \equiv 
 M_q^{-1}P(t_1)^T \big[ \gamma_{1} e^{-i\bm q}  P(t_1) V_{a1} \big].
\label{eq:Va3PC}
\eeq
Here SST stands for {\em Sequential Source Technique} which takes an existing quark propagator as the source for a new quark propagator.

For diagram b and conserved current, 
\fontsize{8}{8}
\beqs  
 \tilde{Q}^{(b, PS)}_{11}(\bm q,t_2)
   ={5\over 9} \kappa^2 \Tr_{s,c} \Bigg[
  \big[P(t_2)\gamma_5 V_{a3}^{(1,PS)}({\bf q})\big]^\dagger \big[ (1-\gamma_{1}) e^{-i\bm q}  U_{1}(t_2,t_2) 
  -(1+\gamma_{1})  U^\dagger_{1}(t_2,t_2)e^{-i\bm q} \big] P(t_2) V_{a2} {\cal W}^T P(t_3)\gamma_5V_{a1}  \Bigg],
\label{eq:Q11b}
\eeqs
where a new inversion is needed for the SST propagator, 
\fontsize{9}{9}
\beq 
V_{a3}^{(1, PS)}(\bm q) \equiv M_q^{-1} \bigg[
   P^T(t_1) (1-\gamma_{1}) e^{-i\bm q} U_{1}(t_1,t_1) P(t_1) V_{a1} 
   -P^T(t_1)(1+\gamma_{1}) U^\dagger_{1}(t_1,t_1)e^{-i\bm q}   P(t_1) V_{a1} \bigg].
\eeq

For diagram c and local current,  we have 
\beq 
\tilde{Q}^{(c, PC)}_{11}(\bm q,t_2)
=-{5\over 9}Z_V^2\kappa^2 \Tr_{s,c} \Bigg[ 
\left[\gamma_1 e^{i\bm q}  P(t_2)\gamma_5 V_{a1}\right]^\dagger  P(t_2) V^{(1,PC)}_{a4}(\bm q) {\cal W}^T P(t_3) \gamma_5  V_{a1}  \Bigg].
\label{eq:cQ44b}
\eeq
 where $V^{(1,PC)}_{a4}$ a SST quark propagator built from $V_{a2}$, 
\beq 
V_{a4}^{(1,PC)}(\bm q) \equiv 
 M_q^{-1}P(t_1)^T \big[ \gamma_{1} e^{i\bm q}  P(t_1) V_{a2}  \big]. 
\label{eq:Va4PC}
\eeq
For diagram c and conserved current, 
\fontsize{8}{8}
\beqs  
 \tilde{Q}^{(c, PS)}_{11}(\bm q,t_2)
   =-{5\over 9} \kappa^2 \Tr_{s,c} \Bigg[
  \big[P(t_2)\gamma_5 V_{a1}({\bf q})\big]^\dagger \big[ (1-\gamma_{1}) e^{-i\bm q}  U_{1}(t_2,t_2) 
  -(1+\gamma_{1})  U^\dagger_{1}(t_2,t_2)e^{-i\bm q} \big] P(t_2) V_{a4}^{(1,PS)} {\cal W}^T P(t_3)\gamma_5V_{a1}  \Bigg],
\label{eq:Q11c}
 \eeqs
where
\normalsize
\beqs 
V_{a4}^{(1,PS)}(\bm q) \equiv 
M_q^{-1} \bigg[P^T(t_1)(1-\gamma_{1}) e^{i\bm q} U_{1}(t_1,t_1)  P(t_1) V_{a2}   
   -  P(t_1)^T (1+\gamma_{1})  U^\dagger_{1}(t_1,t_1) e^{i\bm q} P(t_1) V_{a2} \bigg].
\label{eq:Va4PS1}
\eeqs
 
\end{widetext}

\end{document}